\documentclass[11pt]{article}\textwidth 6.5in\textheight 9in
\usepackage{amssymb}\usepackage[colorlinks]{hyperref}\usepackage{color}
	\usepackage{stmaryrd}\usepackage{mathrsfs}
	\usepackage{graphicx}
	\usepackage{amsmath}
    \usepackage{amssymb}
	\usepackage{tikz}
	\usepackage{braket}
     \usepackage{mathtools}
\usepackage{caption}

\begin{document}
\setlength{\captionmargin}{27pt}

\newcommand\hreff[1]{\href {http://#1} {\small http://#1}}
\newcommand\trm[1]{{\bf\em #1}} \newcommand\emm[1]{{\ensuremath{#1}}}
\newcommand\prf{\paragraph{Proof.}}\newcommand\qed{\hfill\emm\blacksquare}

\newtheorem{thr}{Theorem} 
\newtheorem{lmm}{Lemma}
\newtheorem{cor}{Corollary}
\newtheorem{con}{Conjecture} 
\newtheorem{prp}{Proposition}

\newcommand\QC{\mathbf{QC}} 
\newcommand\C{\mathbf{C}} 

\newtheorem{blk}{Block}
\newtheorem{dff}{Definition}
\newtheorem{asm}{Assumption}
\newtheorem{rmk}{Remark}
\newtheorem{clm}{Claim}
\newtheorem{exm}{Example}
\newtheorem{exc}{Exercise}

\newcommand\Ks{\mathbf{Ks}} 
\newcommand{\ab}{a\!b}
\newcommand{\yx}{y\!x}
\newcommand{\yux}{y\!\underline{x}}

\newcommand\floor[1]{{\lfloor#1\rfloor}}
\newcommand\ceil[1]{{\lceil#1\rceil}}

\renewcommand\do[1]{\overline{\overline{#1}}}

\newcommand{\bmu}{\boldsymbol{\mu}}

\newcommand{\lea}{<^+}
\newcommand{\gea}{>^+}
\newcommand{\eqa}{=^+}

	\newcommand{\bnu}{\boldsymbol{\nu}}

\newcommand{\lel}{<^{\log}}
\newcommand{\gel}{>^{\log}}
\newcommand{\eql}{=^{\log}}

\newcommand{\F}{\mathbf{F}}
\newcommand{\E}{\mathbf{E}}
\newcommand{\lem}{\stackrel{\ast}{<}}
\newcommand{\gem}{\stackrel{\ast}{>}}
\newcommand{\eqm}{\stackrel{\ast}{=}}

\newcommand\edf{{\,\stackrel{\mbox{\tiny def}}=\,}}
\newcommand\edl{{\,\stackrel{\mbox{\tiny def}}\leq\,}}
\newcommand\then{\Rightarrow}

\newcommand{\kpsi}{\ket{\psi}}
\newcommand{\ktheta}{\ket{\theta}}

\newcommand\Ip{\I_\mathrm{Prob}}
\newcommand\uhr{\upharpoonright}
\newcommand\Hg{\mathbf{Hg}}
\newcommand\Hv{\mathbf{Hv}}
\newcommand\Hbvl{\mathbf{Hbvl}}
\newcommand\Hb{\mathbf{Hb}}
\newcommand\Hoc{\mathbf{Hoc}}
\renewcommand\H{\mathbf{H}}
\newcommand\ml{\underline{\mathbf m}}
\newcommand\mup{\overline{\mathbf m}}
\newcommand\UI{\mathcal{I}}
\newcommand{\G}{\mathbf{G}}
\newcommand\km{{\mathbf {km}}}\renewcommand\t{{\mathbf {t}}}
\newcommand\KM{{\mathbf {KM}}}\newcommand\m{{\mathbf {m}}}
\newcommand\md{{\mathbf {m}_{\mathbf{d}}}}\newcommand\mT{{\mathbf {m}_{\mathbf{T}}}}
\newcommand\K{{\mathbf K}} \newcommand\I{{\mathbf I}}

\newcommand\II{\hat{\mathbf I}}
\newcommand\Kd{{\mathbf{Kd}}} \newcommand\KT{{\mathbf{KT}}} 
\renewcommand\d{{\mathbf d}} 
\newcommand\D{{\mathbf D}}
\newcommand\Tr{\mathrm{Tr}}
\newcommand\w{{\mathbf w}}
\newcommand\Cs{\mathbf{Cs}} \newcommand\q{{\mathbf q}}
\newcommand\St{{\mathbf S}}
\newcommand\M{{\mathbf M}}\newcommand\Q{{\mathbf Q}}
\newcommand\ch{{\mathcal H}} \renewcommand\l{\tau}
\newcommand\tb{{\mathbf t}} \renewcommand\L{{\mathbf L}}
\newcommand\bb{{\mathbf {bb}}}\newcommand\Km{{\mathbf {Km}}}
\renewcommand\q{{\mathbf q}}\newcommand\J{{\mathbf J}}
\newcommand\z{\mathbf{z}}
\newcommand\Z{\mathbb{Z}}

\newcommand\Huc{\mathbf{Huc}}
\newcommand\B{\mathbf{bb}}\newcommand\f{\mathbf{f}}
\newcommand\hd{\mathbf{0'}} \newcommand\T{{\mathbf T}}
\newcommand\R{\mathbb{R}}\renewcommand\Q{\mathbb{Q}}
\newcommand\N{\mathbb{N}}\newcommand\BT{\{0,1\}}
\newcommand\W{\mathbb{W}}
\newcommand\dom{\mathrm{Dom}}
\newcommand\FS{\BT^*}\newcommand\IS{\BT^\infty}
\newcommand\FIS{\BT^{*\infty}}
\renewcommand\S{\mathcal{C}}\newcommand\ST{\mathcal{S}}
\newcommand\UM{\nu_0}\newcommand\EN{\mathcal{W}}

\newcommand{\supp}{\mathrm{Supp}}

\newcommand\lenum{\lbrack\!\lbrack}
\newcommand\renum{\rbrack\!\rbrack}

\newcommand\om{\overline{\mu}}
\newcommand\on{\overline{\nu}}
\newcommand\h{\mathbf{h}}
\renewcommand\qed{\hfill\emm\square}
\renewcommand\i{\mathbf{i}}
\newcommand\p{\mathbf{p}}
\renewcommand\q{\mathbf{q}}
\renewcommand\T{\mathbf{T}}

\title{Semi-Classical Subspaces, The No Synchronization Law, and More}

\author {Samuel Epstein\\samepst@jptheorygroup.com}

\maketitle

\begin{abstract}
This paper looks at the intersection of algorithmic information theory and physics, namely quantum mechanics, thermodynamics, and black holes. We discuss theorems which characterize the barrier between the quantum world and the classical realm. The notion of a ``semi-classical subspace'' is introduced. Partial signals and partial information cloning can be obtained on quantum states in semi-classical subspaces. The No Synchronization Law is detailed, which says separate and isolated physical systems evolving over time cannot have algorithmic thermodynamic entropies that are in synch. We look at future work involving the Kolmogorov complexity of black holes.
\end{abstract}

\section{Introduction}
The Turing machine is an abstract computational device, and its universality formulizes the study of algorithms and complexity agnostic to the machine running the programs. This definition leads to the (modern) Church-Turing thesis, which says that  any computational model can be simulated in polynomial time with a probabilistic Turing machine.

However it has become apparent one cannot simply separate computation from the physical model used to perform it. This was made clear with the introduction of the quantum Turing machine (QTM) \cite{DeutschPe85}, on which the famous Shor's Factoring Algorithm \cite{Shor94} can be run efficiently. Using a universal QTM, one can define BQP, which is the class of decision problems solvable in polynomial time by a QTM, with at most 1/3 probability of error \cite{BernsteinVa93}. There is a large group of researchers working on what useful quantum algorithms are in BQP. 

However, an overlooked area of research is the study of the algorithmic entropy level of individual (pure or mixed) quantum states. Simple states such as $\ket{0^n}$ 
should have $O(1)$ entropy, whereas a typical state $\ket{\psi}$ should have $n+O(1)$ entropy. As shown in \cite{EpsteinAPhysics24} there is a wide set of interesting implications and consequences to the study of this area. 

One of the main implications of study into this area is the further characterization of the barrier between the quantum realm and the the classical world. It is easy to see (folklore) there exist no quantum algorithm that can transform a (pure) quantum state into its classical description, up to any accuracy. It has also become apparent that a majority quantum states, when measured by a fixed POVM, will produce either the empty signal or pure white noise (or a combination of the two). Furthermore, in the context of quantum decoherence, most non-pointer states will decohere into a classical probability that is pure white noise. This is a consequence of the fact that quantum states have virtually no self-information, which, as we will see, is shown with two independent information measures. 

This begs the question, why is objective reality so ordered? I suspect it is due to the following reasoning. Due to conservation inequalities (Theorem \ref{thr:infocons}), a quantum operation cannot increase the self-information of a (pure or mixed) quantum state. However a wave function collapse, i.e. measurement, can cause an uptake in algorithmic information of the quantum state. This collapse enables nonnegligible signals from further measurements and also enables partial information cloning. Thus my research leads to the existence of \textit{Quantum}, \textit{Semi-Classical}, and \textit{Classical Subspaces}, ($\mathcal{Q}$, $\mathcal{SC}$, and $\mathcal{C}$), with
$$
\mathcal{Q}\xrightarrow{\textrm{Measurement}}{} \mathcal{SC}\xrightarrow{\textrm{Measurement}}{}\mathcal{C}.
$$
Another application of algorithmic randomness to physics is to determine an objective way to measure the algorithmic content of classical information. As shown with QTMs, algorithms depend on the physical model used to run them. The question is, does the physical model used change the amount information measured in strings. M\"{u}ller's Theorem \cite{Muller07,Muller09} answers this question in the negative; quantum mechanics does not offer any benefit to compressing classical information. The quantitative amount of information in individual strings is not dependent on the physical model used to measure it.

\subsection{The No Synchronization Law}

Classical thermodynamics is the study of substances and changes to their properties such as volume, temperature, and pressure. Substances, such as a gas or a liquid, is modeled as a point in a phase space $\Omega$. There is a volume measure over $\Omega$, denoted by $L$. Boltzmann introduced an entropy measure of points $\omega$ in $\Omega$ using partitions $\{\Gamma_i\}$ of the phase space. The Boltzmann entropy of $\omega\in\Gamma_i$ is $\log L(\Gamma_i)$.

There are several issues with this definition. The Boltzmann entropy depends on the choice of the partition. Another digit of precision will decrease it by around $\log 10$. This issue was addressed in \cite{Gacs94}, which introduced a new entropy measure $\G$, which is the negative logarithm of a universal lower computable $L$-test over $\Omega$. This definition requires a lot of mathematical framework (see Chapter 13 of \cite{EpsteinAPhysics24}), as $\Omega$ can be any computable metric space and $L$ can be any computable (not-necessarily probability) measure over $\Omega$.

However there are many benefits to using $\G$ to model the algorithmic thermodynamic entropy of a state. One property of $\G$ is that it oscillates either under computable discrete ergodic dynamics or under a computable tranformation group. The proof of this uses the same mathematical machinery to prove that outliers must be present in all large enough set of observations.

This paper will discuss a surprising result in \cite{EpsteinAPhysics24} that given two separated and isolated physical systems $\Omega_1$ and $\Omega_2$, under mild conditions, their corresponding algorithmic entropies must be out of sync with respect to ergodic dynamics. Let $L_1$ and $L_2$ be the measures over $\Omega_1$ and $\Omega_2$. Let $\G_1$ and $\G_2$ be the algorithmic entropies corresponding to $\Omega_1$ and $\Omega_2$, respectively. Let $R_1$ and $R_2$ be computable discrete aperiodic measure preserving transformations over $\Omega_1\ni \alpha$ and $\Omega_2\ni\beta$, respectively.\\

\noindent \textbf{No Synchronization Law:} \hspace*{0.5cm}\textit{Under mild conditions, $\sup_{t\in\N} \left|\G_1(R^t_1(\alpha))-\G_2(R^t_2(\beta))\right|=\infty$.}
\section{The Algorithmic Content of Quantum States}
In Algorithmic Information Theory there are two main definitions of the information content of individual strings $x\in\FS$, plain complexity $\C(x)$ and prefx-free complexity $\K(x)$ and they are logarithmically equivalent. In in my manuscript, Algorithmic Physics, \cite{EpsteinAPhysics24}, four definitions are studied. By examining the common traits of these measures, one can distill algorithmic properties of quantum states. 

Let $\mathcal{Q}_n$ be the Hilbert space containing all $n$ qubit quantum states. A matrix, or quantum state, is elementary if all its coefficients are roots of polynomial equations with rational coefficients. One can perform the standard linear algebra algorithms over elementary numbers. Let $D(\sigma,\rho)$ be the trace distance. For positive real function $f$, $\lea f$ means $< f +O(1)$. The four main definitions are as follows:

\subsection{Vit\'{a}nyi Complexity}
Vit\'{a}nyi complexity, \cite{Vitanyi01},  of a pure state $\ket{\psi}$ is equal to the minimum size of a program to a universal Turing machine that outputs an approximation that is an elementary pure state $\ket{\theta}$ of the target state plus a score of their closeness.
$$\Hv(\ket{\psi})=\min_{\textrm{Elementary }\ket{\theta}\in\mathcal{Q}_n}\K(\ket{\theta}|n)-\log |\braket{\psi|\theta}|^2.$$
\subsection{G\'{a}cs Complexity}
G\'{a}cs complexity, \cite{Gacs01}, takes a different approach than Vit\'{a}nyi complexity. The Kolmogorov complexity of a string $x$ is equal to, up to an additive factor, $-\log\m(x)$, where $\m$ is a universal lower-computable semi-measure. Similarly G\'{a}cs complexity is defined using the following universal lower computable semi-density matrix, with
$$\bmu = \sum_{\textrm{Elementary }\ket{\phi}\in\mathcal{Q}_n}\m(\ket{\phi}|n)\,{\ket{\phi}}{\bra{\phi}}.$$ 
The parameter $n$ represents number of qubits used. The G\'{a}cs entropy of a mixed state $\sigma$ is defined by 
$$\Hg(\sigma) = -\log \Tr\bmu\sigma.$$
\subsection{Quantum Unitary Complexity}

Suppose that Alice wants to send a (possibly mixed) $n$ qubit quantum state $\sigma$ to Bob. Alice has access to two channels, a quantum channel and a classical channel. Alice can choose to send $m\leq n$ qubits $\rho$ on the quantum channel and $P$ regular bits $p$ on the classical channel, representing an encoding of unitary operation $V$, and $m$, where  $U(p)= (V,m)$. Bob, upon receiving $\rho$ and $p$, constructs the unitary operation $V$, and then applies it to $\rho$ (tensored with $\ket{0^{n-m}}$) to produce $\sigma'=V(\ket{0^{n-m}}\rho\bra{0^{n-m}}) V^*$. Bob is required to produce $\sigma$ exactly, however there does exist a variant where only an approximation is necessary.

A quantum unitary pair $(V,m)$ consists of two parts, (1) an elementary unitary transform over $\mathcal{Q}_n$, and (2) the number of qbits $m\leq n$ that are connected to the quantum input.
$\mathcal{C}_{n,m}$ be the set of all quantum pairs over $n$ qbits with an input size of $m$. For density matrix $\sigma$, its quantum unitary complexity \cite{EpsteinAPhysics24} is
$$\Huc(\sigma) =\min\{\K(V,m)+m{:}\,(V,m)\in \mathcal{C}_{n,m}, \xi\textrm{ is an $m$ qubit mixed state}, \sigma=V(\ket{0^{n-m}}\xi\bra{0^{n-m}} )V^*\}.$$

\subsection{BvL Complexity}
BvL complexity is the size of the smallest program to a universal QTM $\mathfrak{U}$ that approximates the target (pure or mixed) quantum state. This definition was formulated by \cite{BerthiaumeVaLa01} and the universal QTM $\mathfrak{U}$ used is from \cite{Muller08,Muller09}. The universal QTM has the following property: for every QTM $M$ and mixed state $\sigma$ for which $M(\sigma)$ is defined, there is mixed state $\sigma'$ such that $D\left(\mathfrak{U}(\sigma'),M(\sigma)\right)< \delta,$
		for every $\delta\in\Q_{>0}$ where $\|\sigma'\|\lea \|\sigma\|+\K(M,\delta)$. Furthermore for every QTM $M$ where $M(\sigma,k)$ is defined for all $k\in\N$, there is a $\sigma'$, where $\|\sigma'\|\lea \K(M)$ and $D(\mathfrak{U}(\sigma',k),M(\sigma,2k))< 1/2k$. The BvL complexity of quantum state $\sigma$ is
  $$\Hbvl(\rho)=\min_\sigma \left\{\|\sigma\|:\forall_k, D(\mathfrak{U}\left(\sigma,k\right),\rho)<\frac{1}{k}\right\}.$$
BvL enjoys a special status over the other complexity measures. Whereas $\Hv$, $\Hg$, $\Huc$ can be considered to be algorithmic entropy scores, $\Hbvl$ can be considered to be the minimal amount of quantum mechanical resources needed to construct a quantum state. 
\section{Algorithmic Properties of Quantum States}
Each of the measures above have slightly different properties, but by examining them collectively, one gets a general picture on algorithmic properties of quantum state. The properties examined for the quantum measures are as follows. We use $\H$ to denote a generic measure. 
\begin{itemize}
    \item Subadditivity. $\H(\sigma\otimes\rho)\lea \H(\sigma)+\H(\rho)$.
    \item Monotonicity. $\H(\sigma) \lea \H(\sigma\otimes\rho)$.
    \item Unitary Transform. $\H(U\sigma U^*)\eqa \H(\sigma)\pm\K(U)$.
    \item Addition Inequality. For elementary $\sigma$, $\H(\sigma)+\H(\rho|\langle \sigma\rangle,\ceil{\H(\sigma)})\lea \H(\sigma\otimes\rho)$.
    \item No Cloning Theorem. There exists states $\ket{\psi}^m$ such that $\H(\ket{\psi}^m)$ which have a much higher complexity than $\H(\ket{\psi})$. More formally, $\log{m+2^n-1\choose m}\lea \max_{\ket{\psi}}\H(\ket{\psi}^m)\lea \log{m+2^n-1\choose m}+\K(n,m)$.
     \item M\"{u}ller's Theorem. For $x\in\BT^n$, $\H(\ket{x})$ is equal to $\C(x)$, $\K(x)$, or $\K(x|n)$.
    \item Quantum EL Theorem. A projection $P$ of large rank and large $\min_{\ket{\psi}\in\mathrm{Image}(P)}\H(\ket{\psi})$ will have large mutual information with the halting sequence.
    \item Quantum Levin/Schnorr Theorem. An infinite quantum Martin L\"{o}f sequence \cite{NiesSc19} will have $\H$ incompressible prefixes.
\end{itemize}
\begin{figure}[h!]
	\begin{center}
\begin{tabular}{|c|c|c|c|c|}
\hline
      & $\Hv$ & $\Hg$ & $\Huc$ & $\Hbvl$ \\
     \hline
     Subadditivity & $\checkmark$ & $\checkmark$ & $\checkmark$ & \\
     \hline
     Monotonicity&  & $\checkmark$ & $\checkmark$ & \\
     \hline
     Addition Inequality & & $\checkmark$ & & \\
     \hline
     Unitary Transform& $\checkmark$ & $\checkmark$ & $\checkmark$ & $\checkmark$ \\
     \hline
     No Cloning Theorem & $\checkmark$ & $\checkmark$ & $\checkmark$ & $\checkmark$ \\
     \hline
     M\"{u}ller's Theorem& $\checkmark$ & $\checkmark$ & $\checkmark$ & $\checkmark$ \\
     \hline
     Quantum EL Theorem & $\checkmark$ & $\checkmark$ & $\checkmark$ &  \\
     \hline
     Quantum Levin/Schnorr Theorem & & & $\checkmark$ & \\
     \hline
\end{tabular}
\caption{Properties of four measures of the algorithmic content of quantum states.}
\label{fig:prop}
\end{center}
\end{figure}
Figure \ref{fig:prop} describes the different properties of the quantum complexity measures. Though they have separate properties, there is an overall picture on what it means to quantify the algorithmic content of quantum states. Notice that all measures have a version of M\"{u}ller's theorem. Thus they can be seen as generalizations to the classical measures $\C$ and $\K$. The term $\Hbvl$ takes a special importance because it can be seen as a measure of the amount of quantum mechanical resources needed to approximate or construct a state. This fact, in my opinion, implies M\"{u}ller's Theorem is the greatest accomplishment of the application of algorithmic information theory to physics.
\section{Mutual Information Between Quantum States}
In this section we look at two definitions on the algorithmic mutual information between quantum states and some of their implications.
A positive-semidefinite matrix $A$ is lower computable if there is a program $q$ that outputs a series of elementary positive semidefinite matrices $\{A_i\}$ such that $A_i\leq A_{i+1}$ and $\lim_{i\rightarrow \infty}A_i=A$. Furthermore, we say that $q$ lower computes $A$. Let $\mathcal{C}_{C\otimes D}$ be the set of all lower computable matrices $A\otimes B$, such that $\Tr(A\otimes B)(C\otimes D)\leq 1$. The lower algorithmic probability of a lower computable matrix $\sigma$ is $\ml(\sigma|x) = \sum \{\m(q|x)\,{:}\,q\textrm{ lower computes }\sigma\}$. Let $\mathfrak{C}_{C\otimes D}=\sum_{A\otimes B\in\mathcal{C}_{C\otimes D}}\ml(A\otimes B|n)A\otimes B$.

\begin{dff}
	The mutual information between two quantum states $\sigma$, $\rho$ is defined to be $\I_{\mathrm{d}}(\sigma\,{:}\,\rho)=\log\Tr\mathfrak{C}_{\bmu\otimes\bmu}(\sigma\otimes\rho)$,  where $\bmu$ is the universal lower semi-density matrix.
\end{dff}

The second approach using classical information as a model, where $\I(x:y)=\K(x)+\K(y)-\K(x,y)$.
\begin{dff}
    $\I_\mathrm{g}(\sigma:\rho)=\Hg(\sigma)+\Hg(\rho)-\Hg(\sigma\otimes\rho)$.
\end{dff}

Though the information terms are very different, they share the following key property, in that an overwhelming majority of pure quantum states have no self-information.
\begin{thr} [\cite{EpsteinAPhysics24}]
\label{thr:noinfo}
Let $\Lambda$ be the uniform distribution over all $n$ qubit pure states.
    \begin{itemize}
    \item $\int 2^{\I_{\mathrm{d}}(\ket{\psi}:\ket{\psi})}d\Lambda=O(1)$.
    \item $\int 2^{\I_{\mathrm{g}}(\ket{\psi}:\ket{\psi})}d\Lambda=O(1)$.
    \end{itemize}
\end{thr}
The information terms enjoy conservation inequalities.
\begin{thr}[\cite{EpsteinAPhysics24}]$ $\\
\vspace*{-0.5cm}
	\label{thr:infocons}
 \begin{itemize}
	\item Relativized to elementary quantum operation $\varepsilon$, $\I_{\mathrm{d}}(\varepsilon(\rho):\sigma)\lea \I_{\mathrm{d}}(\rho:\sigma)$. 
 \item Relativized to elementary unitary transform $A$, $\I_{\mathrm{g}}(A\rho A^*:\sigma)\lea \I_{\mathrm{g}}(\rho:\sigma)$.
 \end{itemize}
\end{thr}
\section{Measurements}
In order to better understand the barrier between the quantum and classical realms, we introduce a new information term over probabilities. This is due to the fact that quantum states decohere into probability measures and a POVM will also produce a probability from the measured quantum states. For probabilities $p$, $q$ over strings, their mutual information is defined as follows.
$$
\I_{\mathrm{prob}}(p:q) = \log \sum_{x,y\in\FS}2^{\I(x:y)}p(x)q(y).
$$
This definition obeys conservation inequalities over randomized processing $f:\FS\times\FS\rightarrow \R_{\geq 0}$, where $fp(x)=\sum_z f(x|z)p(z)$.
\begin{thr}[\cite{EpsteinAPhysics24}]
Relativized to random processing $f$,
$$\I_{\mathrm{prob}}(fp:q)\lea\I_{\mathrm{prob}}(p:q).$$
\end{thr}
If $\I_{\mathrm{prob}}(p:p)$ is small then either $p$ represents an empty signal, or white noise, or some combination of the two. Furthermore, no amount of deterministic or randomized processing can increase $p$'s ``signal''.

We recall that a POVM $E$ is a finite or infinite set of positive semi-definite matrices $E_k$ such that $\sum_kE_k=\mathbf{1}$. Given a quantum state $\sigma$, a POVM induces a probability measure $E\sigma(k)=\Tr \sigma E_k$, where $E\sigma(i)$ can be interpreted as the probability of measuring $i$ when $\sigma$ is applied to $E$. The size of POVM $E$ is $|E|$. Remarkably, algorithmic self-information of a quantum state upper bounds the signal produced from a POVM.
\begin{thr}[\cite{EpsteinAPhysics24}] Relativized to elementary POVM $E$,
    $\I_{\mathrm{prob}}(E\sigma:E\sigma)\lea \I_{\mathrm{d}}(\sigma:\sigma)+\log\log |E|$.
\end{thr}
This theorem, combined with Theorem \ref{thr:noinfo}, shows that the measurements of most quantum states produce the empty signal, white noise, or some combination of the two. In fact one can show this directly with the following theorem.
\begin{thr}[\cite{EpsteinAPhysics24}]
Let $\Lambda$ be the uniform distribution over the unit sphere of $\mathcal{Q}_n$.
    Relativized to elementary POVM $E$,
    $$
    \int 2^{\I_{\mathrm{prob}}(E\ket{\psi}:E\ket{\psi})}d\Lambda=O(1).
    $$
\end{thr}

The self information of a quantum state upper bounds the amount of cloneable information it has.
\begin{thr}[\cite{EpsteinAPhysics24}]
    Relativized to elementary POVMs $E$, $F$, and elementary quantum operation $\epsilon$, if $\epsilon(\nu),=\sigma\otimes\rho$, then
    \begin{align*}
    \I_\mathrm{d}(\sigma:\rho) &\lea \I_\mathrm{d}(\nu:\nu),\\
    \I_{\mathrm{prob}}({E\sigma}\,{:}\,{F\rho})&\lea \I_\mathrm{d}(\nu:\nu)+\log\log \max\{|E|,|F|\}.\\
    \end{align*}
\end{thr}

\section{Decoherence}

The following letter of Einstein to Born (April 1954) illustrated the problem of superposition of quantum macrosystems.
\begin{quote}
	\textit{
		Let $\Psi_1$ and $\Psi_2$ be two solutions to the same Schr\"{o}dinger equation\dots When the system is a macrosystem and when $\Psi_1$ and $\Psi_2$ are `narrow
		with respect to position, then in by far the greater number of cases this is no longer true $\Psi_{12}=\Psi_1+\Psi_2$. Narrowness with respect to macrocoordinates is not only independent of the principles of quantum mechanics, but is, moreover, incompatible with them.}
\end{quote}
This letter brings up the astonishing fact that observables on the microscale and absent from everyday experiments. In fact, \textit{quantum decoherence} and \textit{einselection} show that such superpositions are highly fragile and decay exponentially fast. The root cause of this phenomena is caused by interactions between a system and environment. A closed system assumption is a fundamental obstacle to the study of the transition of the quantum domain to the classical domain.

In this light, the setup is a (microscopic) system and (macroscopic) environment. Given joint Hamiltonian dynamics between the system and environment, there are two main consequences.
\begin{enumerate}
	\item The effective disappearance of coherence, the source of quantum interference effects, from the system.
	\item The dynamical definition of preferred ``pointer states'', which are unchanged by the system/environment dynamics.
\end{enumerate}
The phenomena of (1) is called \textit{decoherence} (see \cite{Schlosshauer10} for an extensive overview). The phenomena (2) is called \textit{einselection}, short for Environment INduced Selection \cite{Zurek03}. In Einselection, the system-environment Hamiltonian ``selects'' a set of prefered quasi-classical ``pointer states'' which do not decohere. Einselection explains why we only observe a few ``classical'' quantities such as momentum and positon, and not superpositions of these pointer states.

We begin our explanation with a two state case, which can be generalized to arbitrary number of pointer states. Suppose the system is described by a superposition of two quantum states $\ket{\psi_1}$ and $\ket{\psi_2}$ which for example can be thought of as two localization of two positions $x_1$ and $x_2$ in a double slit experiment. The system/environment interaction results in
\begin{align*}
	\ket{\psi_1}\ket{E_0}&\rightarrow \ket{\psi_1}\ket{E_1}\\
	\ket{\psi_2}\ket{E_0}&\rightarrow \ket{\psi_2}\ket{E_2}.
\end{align*}
So the state of the environment evolves according to the state of the system. Now if the system is in a superposition of $\ket{\psi_1}$ and $\ket{\psi_2}$, we get the dynamics
$$
\frac{1}{\sqrt{2}}(\ket{\psi_1}+\ket{\psi_2})\ket{E_0}\rightarrow \frac{1}{\sqrt{2}}(\ket{\psi_1}\ket{E_1}+\ket{\psi_2}\ket{E_2})
$$
The reduced density matrix of system (with the environment traced out) is
$$
\frac{1}{2}\left(\ket{\psi_1}\bra{\psi_1}+\ket{\psi_2}\bra{\psi_2}+\ket{\psi_1}\bra{\psi_2}\braket{E_2|E_1}+\ket{\psi_2}\bra{\psi_2}\braket{E_1|E_2}\right).
$$
The last two terms correspond to the interference between the state $\ket{\psi_1}$ and $\ket{\psi_2}$. If the environment recorded the position of the particle, then $\ket{E_1}$ and  $\ket{E_2}$ will be approximately orthogonal. In fact, it can be shown that in many dynamics, $\braket{E_1|E_2}\leq e^{-t/\tau}$, where $t$ is the time of the interaction and $\tau$ is a positive constant. In this case
$$
\rho \approx \frac{1}{2}\left(\ket{\psi_1}\bra{\psi_1}+\ket{\psi_2}\bra{\psi_2}\right).
$$
Thus virtually all coherence between the two states $\ket{\psi_1}$ and $\ket{\psi_2}$ is lost. The states $\ket{\psi_1}$ and $\ket{\psi_2}$ are called invariant to the dynamics, and will not undergo decoherence. They are called ``pointer states'' because they induce an apparatus with a pointer mechanism to be orientated at a particular angle. Einselection preserves ``pointer states'' but superpositions of them are fragile and do not survive the dynamics with the system.
\section{Predictability Sieve}
In general, there is not a clear division between pointer and non-pointer states. Instead one can use a score to measure how much of the state has been preserved. The interaction of pointer states with the environment is predictable; they are effectively classical states. However a state that is heavily decohered is unpredictable. Let $\ket{\psi}$ be an initial pure state, and $\rho_{\ket{\psi}}(t)$ be the density matrix of the system state after interacting with the environment for time $t$. The loss of predictability caused by the environment can be measured in the following two measures
\begin{itemize}
	\item $\varsigma^T_{\ket{\psi}}(t) = \Tr \rho_{\ket{\psi}}^2(t).$ 
	\item $\varsigma^S_{\ket{\psi}}(t) = S(\rho_{\ket{\psi}}(t)).$
\end{itemize}
The first measure, uses squared trace of the density matrix whereas the second measure uses von Neumann entropy. The first measure will start at 1 and then decrease proportionately to much much the state decoheres. This is similarly true for the von Neumann entropy predictability sieve, except the measure starts at 0. 

In this section we introduce an algorithmic predictability sieve $\varsigma^A$. Assume a basis of $2^n$ pointer states. Let the system be $\ket{\psi}$, an arbitrary pure state. We consider the limit of interacting with the environment as time approaches infinity. In this idealized case, the decoherence $\ket{\psi}\bra{\psi}$ into a classical probability, with the off-diagonal terms turned to 0. Let $p_{\ket{\psi}}$ be the classical probability that $\ket{\psi}$ decoheres to, with $p_{\ket{\psi}}(i)=\ket{\psi}\bra{\psi}_{ii}$. 

\begin{dff}[Algorithmic Predictability Sieve]
    $\varsigma^A(\ket{\psi}) = \I_{\mathrm{prob}}(p_{\ket{\psi}}:p_{\ket{\psi}}|n)$.
\end{dff}
Thus, $\varsigma^A$ is the self information of the probability measure induced by the diagonal of the density matrix $\ket{\psi}\bra{\psi}$. Note that this self information is relativized to $n$, that is the universal Turing machine $U$ has $n$ on an auxiliary tape. On average, pointer states $\ket{i}$ have high algorithmic predictability. 
	$$ \frac{1}{2^n}\sum_{i=1}^{2^n}\varsigma^A(\ket{i})\eqa n.$$ 
We now show that an overwhelming majority of pure states over the pointer basis decohere into algorithmic white noise. Due to algorithmic conservation inequalities, there is no (even probabilisitic) method of processing this white noise to produce a signal. Thus superpositions of pointer bases will produce garbage that can't be measured. The following statement shows that almost all pure states decohere into algebraic garbage. 	

\begin{thr}
Let $\Lambda$ be the uniform distribution on the unit sphere of an $n$ qubit space. 
$$\int 2^{\varsigma^A(\ket{\psi})}d\Lambda = O(1).$$
\end{thr}

\section{Semi-Classical Subspaces}

As shown in the previous sections, a quantum state's algorithmic self information upper bounds the ``signal'' produced from measurements as well as the amount of information that can be cloned with a quantum operation. In addition, an overwhelming majority of quantum states (pure and mixed) will have negligible self information. The question is how does the ordered classical world emerge from quantum states with no self information?

This section is in support of the general thesis that there are quantum, semi-classical, and classical subspaces. Purely quantum subspaces have quantum states with no self-information, semi-classical subspaces have quantum states with moderate degree of self information, and classical subspaces have quantum states with a large degree of self information. 

Furthermore, measurements are the mechanism for a state to increase its self information and move towards more classical subspaces. To show this, we rely on PVMs and Quantum Bayesianism \cite{ShackBrCa01}.

PVMs is short for projection value measure. A PVM $P=\{\Pi_i\}$ is a collection of projectors $\Pi_i$ with $\sum_i\Pi_i=\mathbf{1}$, and $\Tr \Pi_i\Pi_j=0$ when $i\neq j$. When a measurement occurs, with probability $\bra{\psi}\Pi_i\ket{\psi}$, the value $i$ is measured, and the state collapses to $$\ket{\psi'}=\Pi_i\ket{\psi}{/}\sqrt{\bra{\psi}\Pi_i\ket{\psi}}.$$
Further measurements of $\ket{\psi'}$ by $P$ will always result in the $i$ measurement, so $P\ket{\psi'}(i)=1$.

Quantum Bayesianism \cite{ShackBrCa01} deals with distributions over quantum states. When a PVM is applied to (apriori) distributions over states, a new distribution is created from the probabilistic laws of PVMs. In previous sections, the uniform distribution over pure states $\Lambda$ is used. The purely classical uniform distribution over the basis states has
$$
\frac{1}{2^{n}}\sum_{x\in\BT^n}\I_{\mathrm{d}}(\ket{x}:\ket{x})\gea n.
$$

Let $F$ be a PVM of $2^{n-c}$ projectors, of an $n$ qubit space and let $\Lambda_F$ be the distribution of pure states when $F$ is measured over the uniform distribution $\Lambda$. Thus the support of $\Lambda_F$ represents the semi-classical subspace consisting the $F$-collapsed states from the purely quantum uniform distribution $\Lambda$. Note that if $F$ has too few projectors, it lacks discretionary power to produce a meaningful signal when the states are in distribution $\Lambda_F$. 
\begin{thr}[\cite{EpsteinAPhysics24}]
	\label{thr:pvm}Relativized to elementary PVM $F$,
$n-2c\lea\log \int 2^{\I_{\mathrm{d}}(\ket{\psi}:\ket{\psi})}d\Lambda_F+\log n$.
\end{thr}
This is in contrast to a quantum operation, which cannot increase self information of quantum states. Let $\epsilon$ be an elementary quantum operation and $\Lambda_\epsilon$ be the distribution over quantum states when $\epsilon$ is applied to the uniform distribution $\Lambda$. Due to Theorem \ref{thr:infocons}, we get the following result.
\begin{thr}
    Relativized to elementary quantum operation $\epsilon$,
    $\int 2^{\I_\mathrm{d}(\ket{\psi}:\ket{\psi})}d\Lambda_\epsilon=O(1).$
\end{thr}

\section{No Synchronization Law}
In this section, we detail the No-Synchronization Law. Systems in thermodynamics are represented by\textit{phase spaces} which are assumed to be a computable metric spaces, where every point represents a potential configuration of the system. Computable metric spaces have a dense set of ideal points $x,y\in S$ such that their metric distance $d(x,y)$ is computable. For example, a gas comprised of numerous molecules has a unique dimension for each particle's  positions and momenta (6 in total).
The volume of the space is represented by a computable finite (but not necessarily probabilistic) measure over the phase space. The evolution of the system's state over time is a path through the phase space. By  Louiville's theorem, the transformation is measure preserving.

We discuss some motivation for the definition \textit{algorithmic fine-grained entropy} of a physical state. What's required of this definition is that there is a maximum saturation value where states with this score are maximally entropic. Lower scores represent some degree of non-randomness. To this end, it makes sense to define algorithmic fine-grained entropy as the negative logarithm of a $\mu$ test $t$, where $\mu$ is a computable measure of the phase space. A measure $\mu$ is computable if it is a constructive point in the space of all Borel measures over $S$. In the selection of $t$, it makes sense to choose one that is universal among a class of tests $\mathcal{T}$, where $t\in\mathcal{T}$ and for each $t'\in\mathcal{T}$, there is a $c$ where $ct>t'$. Since tests are lower semi-continuous, it makes sense to choose $t$ to be a universal lower computable test.

It can be shown \cite{HoyrupRo09}, that one can define a universal lower computable $\mu$ test $\t_\mu$ such that $\int \t_\mu d\mu\leq 1$. The algorithmic entropy \cite{Gacs94} of a point $\alpha\in S$ is defined to be $\G_\mu(\alpha)=-\log \t_\mu(\alpha)$. This entropy score has a max value of $\lea \log \mu(S)+\K(\ceil{\log \mu(S)})$ and can take arbitrary negative values, even $-\infty$.

In my opinion, the definition $\G_\mu$ answers how to apply algorithmic randomness to thermodynamics. The term is very easy to work with, as there are many properties that can be proved with $\t_\mu$ which then can be directly translated into properties about $\G_\mu$. We detail one such property, that two separate, isolated physical systems that are typical must be out of sync.
\begin{thr}[\cite{EpsteinAPhysics24}]
Let $(\mathcal{X}\times \mathcal{Y},\mu\times \nu)$ be a computable product measure space.  Let $R^t_\mathcal{X}$ and $R^t_\mathcal{Y}$ be computable aperiodic measure preserving transformations over $\mathcal{X}$ and $\mathcal{Y}$ respectively.  Let $(\alpha,\beta)\in \mathcal{X}\times \mathcal{Y}$. If $\G_{\mu\times\nu}(\alpha,\beta)>-\infty$ and $(\alpha,\beta)$ has finite mutual information with the halting sequence then $\sup_{t\in\N}|\G_\mu(R^t_\mathcal{X}(\alpha))-\G_\nu(R^t_\mathcal{Y}(\beta))|=\infty$.
\end{thr}
The notion of ``finite mutual information with the halting sequence'' merits some further explanation. For $(\alpha,\beta)\in \mathcal{X}\times\mathcal{Y}$, let $\Pi\subset\IS$ be the set of all encoded infinite sequences consisting of ideal points that converge exponentially quickly to $(\alpha,\beta)$. The point $(\alpha,\beta)$ has finite mutual information with the halting sequence $\mathcal{H}$ if there is a sequence $\gamma \in \Pi$ such that $\I(\gamma:\mathcal{H})<\infty$, where $\I$ is any suitable information term between infinite sequences (say, the one from \cite{Levin74}). Due to the Independence Postulate, \cite{Levin84,Levin13}, constructs with high mutual information with the halting sequence, are exotic and cannot be found in the physical world. Thus separate systems with synchronized algorithmic thermodynamic entropies can be considered unphysical.

\section{Looking Forward: Black Holes}
An overwhelming majority of research in Algorithmic Physics has been in quantum mechanics or thermodynamics. However this represents just the starting point for the investigation of the intersection of physics with algorithmic information theory. For example, Brown and Susskind introduced \cite{BrownSu18} the Kolmogorov complexity of black holes. They did this by associating black holes with the $SU(n)$ space, then partitioning this space into cells, assigning each cells a number $c$, then defining the Kolmogorov complexity of all unitary transforms in that cell with $\K(c)$. Each cell can be seen as a node of a graph $G$ and two vertices $v_1$, $v_2$ share an edge if there are to unitary transforms $U_1\in v_1$ and $U_2\in v_2$ such that $U_2 = gU_1$, where $g$ is a simple gate. The evolution of the complexity of a black hole can be seen as the evolution of the Kolmogorov complexity of the nodes of a random walk over $G$.

Brown and Susskind \cite{BrownSu18} conjectured that the complexity of a black hole will increase linearly for an exponential amount of time before reaching a maximum and fluctuating continuously after that. This would match the volume of the Einstein-Rosen bridge between two entangled black holes, and is called the \textit{Complexity/Volume Correspondence}. This conjecture was proven to be true for quantum circuit complexity \cite{HaferkampFaKoEiHa22}, but it remains open for Kolmogorov complexity. In \cite{EpsteinAPhysics24}, it was shown if $G$ is an expander graph, then linear growth of Kolmogorov complexity of black holes must occur (from \cite{HaferkampFaKoEiHa22}, one can assume proving expander properties for $G$ will land someone a Nature publication!).


\newcommand{\etalchar}[1]{$^{#1}$}

\end{document}